*Laser writable high-K dielectric for van der Waals nano-electronics*

N. Peimyoo,[1] M. D. Barnes,[1] J. D. Mehew,[1] A. De Sanctis,[1] I. Amit,[1] J. Escolar,[1] K. Anastasiou,[1] A. P. Rooney,[2] S. J. Haigh,[2] S. Russo,[1] M. F. Craciun,[1] and F. Withers[1*]

[1]Centre for Graphene Science, College of Engineering, Mathematics and Physical Sciences, University of Exeter, Exeter EX4 4QF, United Kingdom.

[2]School of Materials, University of Manchester, Oxford Road, Manchester M13 9PL, United Kingdom.

*Corresponding author email: f.withers2@exeter.ac.uk

**Abstract**

Like silicon-based semiconductor devices, van der Waals heterostructures will require integration with high-K oxides. This is needed to achieve suitable voltage scaling, improved performance as well as allowing for added functionalities. Unfortunately, commonly used high-$k$ oxide deposition methods are not directly compatible with 2D materials. Here we demonstrate a method to embed a multi-functional few nm thick high-$k$ oxide within van der Waals devices without degrading the properties of the neighboring 2D materials. This is achieved by in-situ laser oxidation of embedded few layer HfS$_2$ crystals. The resultant oxide is found to be in the amorphous phase with a dielectric constant of $k \sim 15$ and break-down electric fields in the range of 0.5-0.6 Vnm$^{-1}$. This transformation allows for the creation of a variety of fundamental nano-electronic and opto-electronic devices including, flexible Schottky barrier field effect transistors, dual gated graphene transistors as well as vertical light emitting and detecting tunneling transistors. Furthermore, upon dielectric break-down, electrically conductive filaments are formed. This filamentation process can be used to electrically contact encapsulated conductive materials. Careful control of the filamentation process also allows for reversible switching between two resistance states. This allows for the creation of resistive switching random access memories (ReRAMs). We believe that this method of embedding a high-$k$ oxide within complex van der Waals heterostructures could play an important role in future flexible multi-functional van der Waals devices.

## Introduction

The high quality of the native oxide which can be grown on the surface of silicon has underpinned the wide success of modern micro and nano-electronics. In recent years, high-$k$ dielectrics such as $HfO_2$ have been adopted in order to reduce the dimensions of nano-electronic components and boost their performance (*1*). Recent work has shown similar native oxides in 2D materials such as $HfSe_2$, $ZrSe_2$ (*2*), $TaS_2$ (*3*) and $TaSe_2$ (*4*). However, use of these oxides embedded within van der Waals (vdW) heterostructures has not been shown.
In comparison to silicon, vdW heterostructure devices are likely to play an important role in future electronic device applications (*5*). With a rapidly growing family of layered two-dimensional (2D) materials (*6*), the multitude of possible heterostructure combinations available will allow for device designs with unprecedented functionalities and improved performance (*2*). To date, many such vdW heterostructure devices have been shown, such as: vertical tunneling transistors (*7*) with negative differential resistance (*8*), light emitting quantum wells (*9, 10*), photovoltaics (*11-16*) and memory devices (*17*).
Contrary to the conventional molecular beam epitaxy (MBE) growth of semiconductor devices, vdW heterostructures make it possible to produce atomically sharp interfaces between different materials (i.e. semiconductors, insulators, semimetals, etc.) without concerns for their inter-compatibility during fabrication. Most significantly, the absence of dangling bonds on the surface of atomically thin materials allows for the creation of atomically sharp interfaces, eliminating the problem of inter-diffusion known to impose severe limitations on the down scaling of devices fabricated by standard semiconductors. So far, the state-of-the-art vdW devices studied experimentally rely on the use of high purity hexagonal boron nitride (hBN) as a gate dielectric, a tunnel barrier or as a high-quality substrate material (*18*). Such high-quality hBN crystals are not widespread and scalable CVD versions typically contain impurities which lead to significant leakage current in transistor devices (*19, 20*). Furthermore, the dielectric constant of hBN ($k \cong 4$) is comparable to that of $SiO_2$ ($k = 3.9$), thus limiting the scaling down in vdW nano-electronics (*21*). Common deposition techniques used for $SiO_2$ and $HfO_2$ are not directly compatible with 2D materials (*22, 23*). In general, such methods tend to damage or modify the electronic properties of the underlying 2D crystal (*24*), especially when the 2D material is thinned to single unit cell thickness. Other options include exploring atomically flat layered oxides such as mica or $V_2O_5$ and assembling them layer by layer. However, these dielectrics also result in a significant level of charge transfer to neighboring 2D materials, large hysteresis in field effect devices and significant reduction of the mobility (*25*). Therefore, the search for alternative dielectrics or novel technologies, compatible with 2D materials, which give good interface quality and with high-$k$, is needed.

In this work we demonstrate a novel route to embed ultra-thin $HfO_x$ in vdW heterostructures using selective photo-oxidation of $HfS_2$. $HfS_2$ is a layered semiconductor with an indirect bandgap of 2.85 eV in its bulk form (*26-28*) and has comparable surface roughness to other 2D crystals after exfoliation (see supplementary materials for atomic force microscopy). We found that the photo-oxidation process can be enabled using laser light even when the $HfS_2$ is embedded within complex heterostructures and under metallic contacts. This fabrication technique eliminates the need for invasive sputtering or ALD methods (*29*). We demonstrate that the photo-induced $HfO_x$ has a dielectric constant $k \cong 15$ and that this dielectric can be incorporated into four classes of devices enabling different applications: flexible field-effect transistors (FETs), resistive switching memories (ReRAMs), vertical light emitters and photodetectors.

## Results
### Photo-oxidation of $HfS_2$ in vdW heterostructures
The procedure used to fabricate our devices is illustrated in Figure 1A. Heterostructures are assembled using dry transfer of micro-mechanically exfoliated 2D crystals (*25, 30, 31*). A few layer flake of $HfS_2$ crystal is placed in the stack where the dielectric is required (Figure 1A, left). In more complex structures additional layers are subsequently transferred (See section S1 for details). Once the device has been produced and the contacts defined by electron-beam lithography, the desired region of oxide is selectively irradiated using visible laser light (Figure 1A, center and right, see methods). It has been shown that upon laser irradiation, thin $HfS_2$ undergoes an oxidation reaction due to the charge transfer between the semiconductor and the water redox couple present on its surface (*32*) and it converts into an oxide of hafnium.

By comparing laser irradiation effects of HfS$_2$ in vacuum and in atmosphere we show that this process relies on the presence of atmospheric water and/or oxygen, Figure 1E. Furthermore, the oxidation process is still found to occur even when the HfS$_2$ is sandwiched between neighboring 2D materials. Figure 1E shows a region of HfS$_2$ stacked within graphene and hBN after laser assisted oxidation (red hatched region). The mechanism likely involves migration of interfacial water between the graphene-HfS$_2$/HfO$_x$ interface to the reaction site which is being irradiated. It should be noted that upon exfoliation the surface layer of the HfS$_2$ will naturally oxidize in the 10-15 min before encapsulation which could allow for diffusion of atmospheric water between the graphene and more hydrophilic HfO$_x$ surface. Similar diffusion effects have previously been observed for graphene on SiO$_2$ (*33*). Figure 1B shows a high resolution scanning transmission electron microscopy (HR STEM) image (*34, 35*) of the cross-section of a Graphene/HfO$_x$/Graphene device, where the few- layer top and bottom graphene electrodes are still clearly visible whilst the long-range crystal order of HfS$_2$ is lost and the resultant material appears in an amorphous phase. Energy dispersive x-ray spectroscopy (EDX) analysis confirms that the only species present in this phase are hafnium and oxygen, with only low levels of sulphur left after laser irradiation, as shown in Figure 1B. We find that graphene encapsulated in laser written HfO$_x$ (red curve) shows only a slight reduction in field effect mobility (~1500-5000 cm$^2$V$^{-1}$s$^{-1}$ at carrier concentration of ~ $10^{11}$ cm$^{-2}$) compared to graphene on our source of hBN (blue curve) and only a small level of n-type doping, Figure 1F. Inset of Figure 1F shows Raman spectrum of graphene encapsulated by HfO$_x$ after photo-oxidation, in which a negligible D peak is seen. This indicates that graphene is not significantly structurally damaged by the laser irradiation process.

We characterized the insulating properties of the laser-written HfO$_x$ by fabricating a MoS$_2$ field-effect transistor (FET) with an 8 nm HfS$_2$ flake separating a graphene gate electrode and the Cr/Au contacts, as shown in Figure 1C. After laser exposure, the transparency of the HfS$_2$ film increases significantly, indicating an increase in the band-gap from 2.85 $eV$ consistent with the formation of an oxide (E$_g$~ 5.5 $eV$, expected for HfO$_x$ (*36*). Vertical electron transport through the oxide further supports the transformation to the oxide and indicates that oxidation not only occurs under the flakes of 2D materials, but also under thick Au contacts ($d = 60\ nm$) facilitated by diffraction of the laser beam around the ~ μm wide contacts. Figure 1D shows the $I_{sd} - V_{sd}$ characteristics for such device. Before oxidation the I$_{sd}$-V$_{sd}$ shows the typical non-linear behavior expected for tunneling through a series of semiconducting materials (*37*), with a low-bias vertical resistance $R{\sim}20 \cdot 10^6\ \Omega\ \mu m^2$. After oxidation, the resistance around V$_{sd}$ = 0 V increases to $R{\sim}10^{11}\ \Omega\ \mu m^2$ consistent with the increase in the tunneling barrier height.

The break-down voltage of the laser-written oxide was measured using a Graphite/HfO$_x$/Cr (5 $nm$) /Au (60 $nm$) vertical electron tunneling device, schematically shown in the inset of Figure 2A. Tunneling current can be measured when a source drain bias is applied across the vertical junction, as shown in Figure 2A. The tunnel current is found to increase exponentially until an electric field of $E_{BD}{\sim}0.5 - 0.6\ V/nm$ is applied, at which point the current discontinuously increases to the compliance level of the voltage source-meter, Figure 2B. This break-down field is comparable to that of SiO$_2$ and hBN 0.6-2.5 V/nm and 1 V/nm respectively (*38, 39*). The tunnel current is well fit by the Fowler-Nordheim tunneling model, inset of Figure 2A. We are able to estimate the barrier height value of $\Phi_B{\sim}1.15$ eV (*40*). This value is smaller than expected for a graphene-HfO$_2$ barrier $\Phi_B{\sim}1.78$ eV, likely related to the non-stoichiometry of the amorphous oxide and finite impurity content leading to an impurity band forming below the conduction band edge. Scaling of the tunnel conductivity with oxide thickness was found to be unreliable with thin oxides d < 3 nm displaying significantly lower than expected resistance (~$10^{6-7}\ \Omega\ \mu m^2$) while oxides of thickness d > 10 nm show similar resistance to 5 nm thick oxides (~$10^{11-12}\ \Omega\ \mu m^2$) (Resistances normalized to device area). This can be explained as follows: In thin flakes there is a higher chance for electrical pin-holes caused by impurities or defects which shunt the current away from high resistance paths. Whilst thicker flakes do not fully oxidize for the same irradiation energy (which was kept constant in this work) leading to higher than expected conductivity. Optimal thickness for uniform oxidation was found to be 4-8 nm. We expect the oxide quality could easily be improved by optimizing for laser excitation energy, excitation power and laser spot dwell time during the writing procedure. Current tunneling through thin HfO$_x$ dielectric was also measured using conductive atomic force microscopy (CAFM), see supplementary materials, Figure S5.

To better understand the dielectric properties of the laser-written HfO$_x$, we fabricated dual-gated graphene field-effect transistors (FETs). An optical micrograph of a FET constructed on a Si/SiO$_2$ (285nm) substrate from a stack of bi-layer graphene/HfO$_x$ (7 nm) and Cr/Au contacts is shown in the inset of Figure 3B. The metal contacts are placed directly on the bilayer graphene (contacts 1, 2 and 11) and on top of the HfO$_x$ (contacts 3-10). To form a contact between the top Cr/Au metal lead and the graphene underneath the HfO$_x$ we rely on the formation of a stable conductive filament produced by the intentional breakdown of the dielectric. In this way we can use, for example, contacts 7 and 8 in the inset of Figure 3B as source and voltage probes and contacts 9,10, and 11 as drain and voltage probes, while the other metal leads (3-6 and 8) are used as top-gates. The I$_{sd}$-V$_{sd}$ characteristics showing stable filament formation are shown in Figure 3A, where the red curve shows initial dielectric break-down at a vertical electric field of $\sim 0.5 \ V/nm$. Further cycling of the source-drain bias with increasing current compliance leads to stable non-reversible filament formation which allow for direct contacting of the underlying graphene channel. Typical contact resistances of $\sim 5.5 \ k\Omega$ are achieved after filamentation (as the area of conductive filament is unknown, we cannot estimate the resistivity in this case). Back-gate (SiO$_2$) sweeps of the resistance show the bilayer graphene to be heavily p-type doped with the charge neutrality point (CNP) lying at $V_{CNP} \sim 80 \ V$. Such p-type doping levels are attributed to the oxygen plasma cleaning of the Si-SiO$_2$ substrate, used to promote the adhesion of graphene prior to exfoliation. Similar hBN-Graphene-HfO$_x$ stacks show negligible doping compared to graphene on hBN (Figure 1F and supplementary Figure S6).

Figure 3C shows a four-terminal top gate – back gate contour plot of the 4-point channel resistance between contacts 7 and 9 (contacted through filamentation) with contact 8 serving as the top gate electrode. From the slope of the neutrality point, $dV_{tg}/dV_{bg}$, and the thickness of the oxide (determined from AFM data), we extract the dielectric constant of the HfO$_x$ material to be $k \sim 15 \pm 1$. This value is similar to literature values for amorphous HfO$_x$ (*41, 42*). Therefore, having confirmed that the dielectric properties of our laser-written HfO$_x$ are comparable to those of sputtered HfO$_x$ films we turn our attention to its implementation in electronic devices.

**Laser-written HfO$_x$ as high-*k* dielectric for 2D field-effect transistors**

A drawback of graphene FETs is the absence of a bandgap which prevents the use of this single layer of carbon atoms in practical field effect transistors, where a suitable I$_{on}$/I$_{off}$ ratio is required. In contrast, few-layer transition metals dichalcogenide (TMDC) are semiconductors and yet atomically thin (*43*), therefore ideally suited for transistor applications. We explored the fabrication of TMDC-FETs using the laser defined ultra-thin and self-assembled high-*k* HfO$_x$ on Si/SiO$_2$ and flexible PET substrates. We first study the performance of such devices on a rigid Si/SiO$_2$ substrate, as schematically illustrated in Figure 4A, inset. Applying a voltage to the graphene electrode (V$_{bg}$) allows us to modulate the carrier injection into the MoS$_2$ channel. The two-terminal gate dependence of the source-drain channel current (I$_{sd}$) for a few-layer MoS$_2$ FET at different source-drain bias voltages (V$_{sd}$) is shown in Figure 4A and B. We find that such devices have turn-on voltages $V_g \sim -0.4 \ V$ with $I_{on}/I_{off} \sim 10^4$ and subthreshold swings as low as $100 \ mV/dec$. Negligible levels of hysteresis are observed in our device, as shown in Figure 4B for a sweep rate 0.3V/min and V$_b$ = 10 mV. Higher levels of hysteresis are typically seen for TMDC FETs on SiO$_2$ substrates due to the presence of water and oxygen which act as electric field dependent dopants (*44-46*). Field-effect mobilities in the linear region for our MoS$_2$ FETs are found to be $\mu \sim 1-2 \ cm^2 V^{-1} s^{-1}$, comparable to MoS$_2$ FETs on SiO$_2$ (*47, 48*). The absence of significant hysteresis highlights the high quality and low impurity content of our dielectric. To further understand the level of charge traps in our HfO$_x$, we systematically investigate the hysteretic behavior of graphene and MoS$_2$ devices in different dielectric environments (see supporting information, Figures S6 and S7).

To test the suitability of our HfO$_x$ for flexible applications we prepared a multi-layer MoS$_2$ FET on a 0.5 mm thick PET substrate and subjected it to uniaxial strains of up to 1.6 % in a custom-made bending rig (Figure 4D inset). The I$_{sd}$-V$_g$ sweeps are shown in Figure 4C, where no significant change in the device performance is observed after applying increasing levels of strain. Such devices operate over many bending cycles without degradation as shown in Figure 4D, with a gate leakage current at V$_g$ = 1.5 V less than 40 pA and a small variation in the I$_{sd}$ at a bias voltage of V$_{sd}$ = 20 mV.

**Resistive switching memory devices**
The formation of conducting filaments illustrated in Figure 2, allows for switching between two resistance states, creating a device known as resistive switching random access memory element (ReRAM) (*49*). ReRAM devices represent a promising emerging memory technology with several advantages over conventional technologies including increased speed, endurance and device density. Of several groups of materials that show resistive switching, transition metal oxides including HfO$_x$ are promising candidates (*50*). More recently, such devices based on two-dimensional materials are beginning to attract attention owing to high mechanical flexibility, reduced power consumption and potential for high density memory devices based on stacks of vdW heterostructures (*51*).

Figure 5 shows representative device characteristics for a typical resistive switching element based on photo-oxidized few-layer HfS$_2$. Our devices consist of an Au top electrode with either titanium or chromium used as an adhesion layer deposited on top of the HfS$_2$-graphite heterostructure (see inset Figure 5A). Following photo-oxidation of the HfS$_2$, the device is subjected to repeated current-voltage sweeps, where the top metal electrode is voltage biased with respect to the bottom graphite electrode. During the initial voltage sweeps the current compliance and bias voltage are incrementally increased until stable and repeatable resistance cycling is achieved (it is important to note that increasing the current compliance and V$_b$ further will lead to non-reversible conductive filaments). Figure 5a shows a subsequent switching loop after initial breakdown. At +1 V an abrupt increase in current is observed as the device switches from the high resistance state (HRS) to a low resistance state (LRS), known as the SET process. The device maintains its LRS as the polarity is reversed and swept down to -1 V, at which a reduction in current for increasing negative voltage is observed, as the device switches back to the HRS, known as the RESET process. The use of thin flakes allows for low voltage operation with the SET/RESET voltages around |V$_{sd}$| ~ 1 V. The memory window of devices measured here (R$_{HRS}$ / R$_{LRS}$) varies from ~ 5 up to 10$^4$, with the larger values observed for Au/Ti top electrodes (see supplementary material, Figure S8). Figure 5B shows similar current-voltage behavior for the 1$^{st}$ and 100$^{th}$ cycle. The results of repeated cycling are shown in Figure 5C in which R$_{HRS}$/R$_{LRS}$ (with both resistance values extracted at V$_{sd}$ = 100 mV), shows little variation over 100 cycles. Finally, we investigate the long-term stability of this ReRAM device, Figure 5D, and find that the resistance levels, measured for V$_{sd}$ = 250 mV, are consistent and well defined for over 10$^4$ seconds. We note that resistive switching in devices utilizing graphene for both top and bottom electrodes was unreliable, and we postulate that electrode material asymmetry is crucial for reliable device performance. Such bipolar switching is consistent with the formation and rupture of conducting filaments, however further studies are required to optimize device performance and to better understand the role played by disorder, oxide thickness and contact chemistry.

**Optoelectronic devices**
As discussed above, vertical electron transport in HfO$_x$ formed from thin 1-3 layer parent HfS$_2$ crystals allowed for much higher tunneling currents. Whilst such high leakage currents are detrimental in transistor applications other device types require higher electron transparency and higher injection rates. As such we made use of this property of thinner oxide flakes to realise light emitting and detecting tunneling transistors.

Such vertically stacked heterostructures of two-dimensional materials provide a framework for the creation of large-area, yet atomically thin and flexible optoelectronic devices with photodetectors (*11, 13, 52*) and light-emitting diodes (*9, 10, 53*). So far only hBN tunnel barriers have been demonstrated, however other wide gap material oxides have not been explored when combined with vdW heterostructures. Here we demonstrate the use of ultra-thin $HfO_x$ tunnel barriers in vertical light emitting tunneling transistor device geometries.

Figure 6A shows a current-voltage curve of a $HfO_x$ single-quantum well device (SQW) formed by the encapsulation of $MoS_2$ in 1-2 nm of $HfO_x$. Applying a bias voltage between the top and bottom graphene electrodes ($G_t$ and $G_b$) allows a current to tunnel through the thin $HfO_x$ layers and into the $MoS_2$. There is a negligible temperature dependence of the measured source-drain current, indicating a tunneling mechanism rather than transport through low energy impurity states, see Supplementary Materials, Figure S9). As we increase the bias voltage from zero the current increases non-linearly. Outside of a low-bias regime ($|V_{sd}| > 1$ V) we observe an increase in the current due to tunneling into the conduction band of $MoS_2$. In addition, an asymmetry between the current at positive and negative bias voltage is observed which is likely due to both a variation in doping between $G_t$ and $G_b$ and a different thickness of the top and bottom $HfO_x$. This behavior is similar to previous work using hBN tunnel barriers (*9*).

To determine the active area of the heterostructure we use scanning photocurrent microscopy (SPCM) whereby a laser beam is rastered across the device whilst photocurrent is acquired simultaneously, see Methods. Figure 6B shows that under a moderate bias ($V_{sd} = -1$ V) the photocurrent is predominately localized to regions of overlap between the top and bottom graphene flakes, each outlined in light green. Photoexcited carriers in $MoS_2$ (red outline) are separated by the graphene electrodes due to the applied vertical electric field. Away from this region the photocurrent ($I_{pc}$) drops from > 65 nA to < 10 nA. In Figure 6C we measure a reduction in the magnitude of the photocurrent as we increase the light modulation frequency. By normalizing this to the value of the photocurrent at low frequencies $I_{pc}^0$ we can ascertain the -3dB bandwidth of the device, which we find to be $f_{-3dB} = 40$ kHz. From this we can estimate the rise time using $\tau_r = 0.35/f_{-3dB} \sim 8.8$ microseconds (μs) which is in good agreement with our analysis of the temporal response of the photocurrent, see supplementary materials. The insert of Figure 6C shows multiple iterations of the photocurrent obtained at 1.8 kHz. The measured response time is $10^3$-$10^6$ times faster than typical planar $MoS_2$ photodetectors (*54*), a result arising from the use of a vertical, as opposed to lateral, contact geometry. The small electrode separation ~6 nm and large electric fields ~0.1-0.2 V/nm minimize the transit time of the photoexcited carriers. Hence, these vertical heterostructures of $MoS_2$ encapsulated in $HfO_x$ are a promising high-speed light-detection architecture.

As the bias voltage is further increased, the quasi fermi-levels of the graphene electrodes allow for simultaneous injection of electrons into the conduction band of $MoS_2$ and holes into the valence band. The carrier confinement set by the $HfO_x$ tunnel barriers allows for exciton formation in the $MoS_2$. The subsequent decay of those excitons leads to light emission at the excitonic gap of $MoS_2$. Figure 6D shows the electroluminescence (EL) intensity map as a function of photon energy and bias voltage, where the main EL band appears at 1.78 eV. Line plots of the EL spectra at 0.1 V increments are shown in supplementary materials.

In brief, EL is not observed when $V_{sd} > -2$ V. Only upon reducing the bias voltage below -2V can EL be detected with a more intense signal recorded by increasing $|V_{sd}|$. The emergence of EL at -2 V corresponds well with the single particle band-gap of mono-layer $MoS_2$ (*55, 56*) whilst the negative threshold voltage can be attributed to the asymmetric device structure.

Figure 6E shows a false-color CCD image of the EL overlaid on a monochrome image of the device at applied bias voltage of -2.5 V. The EL is localized to the active area of the device previously identified in Figure 6B through photo-current mapping. To further understand the emission, normalized EL and photoluminescence (PL) spectra are shown in Figure 6F. The main PL emission peak is assigned to the A exciton seen at an energy of 1.8 eV. The energy of the main EL band redshifts from that of PL by 53 meV. Typically, the exfoliated monolayer $MoS_2$ is n-doped, which favors the formation of negatively-charged excitons (*57*), which have a lower emission energy than that of the neutral exciton by ~30 meV. Therefore, we attribute the main feature in electroluminescence spectra at 1.78 eV to the radiative recombination of the charged exciton. Moreover, the dissociation energy (i.e. energy shift referring that of neutral exciton) of charged exciton is proportional to the doping con-centration (*57*). Therefore, it is likely the large energy difference between electroluminescence and photoluminescence is an indication of high doping in monolayer $MoS_2$, which is due to doping of the as-exfoliated natural $MoS_2$ fakes and extra charge transfer from $HfO_x$.

## Conclusion

In conclusion, we show that ultra-thin few layer HfS$_2$ can be incorporated into a variety of vdW heterostructures and selectively transformed into an amorphous high-$k$ oxide using laser irradiation. Contrary to sputtering or ALD, the use of photo-oxidized HfS$_2$ allows for clean interfaces, without damaging the underlying 2D materials. We demonstrate that the laser-written HfO$_x$ has a dielectric constant $k \sim 15$ and a breakdown field of $\sim 0.5 - 0.6\ V/nm$. These properties allow us to demonstrate several promising high-quality vdW heterostructure devices using this oxide: (1) ReRAM memory elements which operate in the $\sim 1\ V$ voltage limit; (2) flexible TMDC-FETs with $I_{on}/I_{off} > 10^4$, subthreshold swings of $100\ mV/dec$ and good resilience to bending cycles; (3) optoelectronic devices based on quantum-well architectures, which can emit and detect light in the same device, with EL intensities and drive voltages comparable to devices with hBN barriers and photodetection response times up to $10^6$ times faster than equivalent planar MoS$_2$ devices. Moreover, the high-$k$ dielectric constant, the compatibility with 2D materials and the ease of laser-writing techniques (*58*) will allow for significant scaling improvements and greater device functionality, which we predict to be an important feature for future flexible semi-transparent van der Waals nano-electronics.

## Materials and Methods

### Device fabrication

Devices were fabricated using standard mechanical exfoliation of bulk crystals and dry transfer methods utilized to form the heterostructures (see supplementary materials for details). Following heterostructure production the contacts were structured using either optical or electron beam lithography. Followed by thermal evaporation of Cr/Au (5/60nm) electrodes.

After vdW assembly, photo-oxidation of the HfS$_2$ layer was performed by rastering either UV ($\lambda_{in} = 375\ nm$) or visible ($\lambda_{in} = 473\ nm$) laser light focused to a diffraction-limited spot in a custom-built setup (*59*). A typical energy density of $53\ mJ/\mu m^2$ was used for exposures lasting 1-2 seconds per point of the HfS$_2$ layer. The focused spot-size was $d_s = 264\ nm$ for the UV laser and $d_s = 445\ nm$ for the visible wavelength.

### Materials characterization

<u>STEM imaging.</u> A cross sectional specimen for high-resolution scanning transmission-electron microscopy (HR STEM) was prepared in a FEI Dual Beam Nova 600i instrument incorporating a focused ion beam (FIB) and a scanning electron microscope (SEM) in the same chamber. Using 30 kV ion milling, platinum deposition and lift-out with a micromanipulator, a thin cross section of material was secured on an Omniprobe TEM grid and thinned down to electron transparency with low energy ions. HR STEM images were acquired using a probe side aberration-corrected FEI Titan G2 $80- 200$ kV with an X-FEG electron source. Bright-field (BF) and high angle annular dark-field (HAADF) imaging were performed at 200 kV using a probe convergence angle of 21 mrad, a HAADF inner angle of 48 mrad and a probe current of ~80 pA. The lamellae were aligned with the basal planes parallel to the incident electron probe. Correct identification of each atomic layer within bright-field and HAADF images was achieved by elemental analysis with energy dispersive X-ray (EDX) spectrum imaging.

<u>Atomic force microscopy</u> was performed using a Bruker Innova system operating in the tapping mode, to ensure minimal damage to the sample's surface. The tips used were Nanosensors PPP-NCHR, which have a radius of curvature smaller than 10 nm and operate in a nominal frequency of 330 kHz.

**Electrical measurements**

$I_{sd}$-$V_{sd}$ were collected using a Keithley 2400 voltage/current source meter. Electrical characterization of graphene and TMDC FET's were performed using standard low noise AC Lock-in techniques using a Signal Recovery 7225 lock-in amplifier and a Keithley 2400 source-meter providing the gate voltage.

All electrical transport measurements were performed in either a vacuum of $10^{-3}$ mbar or a dry helium atmosphere at room temperature unless otherwise stated. The flexible $MoS_2$ FET produced on PET was measured in ambient conditions.

**Optoelectronic characterization**

Optoelectronic measurements were performed using a custom-built setup (*59*). Photocurrent measurements were performed using a continuous-wave laser ($\lambda_{in} = 514\ nm$, $P = 15\ W/cm^2$) rastered on the devices to produce spatial maps of the photo-response. The electrical signal was acquired by a DL Instruments Model 1211 current amplifier connected to a Signal Recovery model 7124 digital signal processing lock-in amplifier. The frequency modulation of the lasers was 73.87 Hz. Electroluminescence and photoluminescence measurements were performed in the same setup using a Princeton Instruments SP2500i spectrometer and PIXIS400 camera. All measurements were performed at room temperature under vacuum ($P = 10^{-5}$ mBar).

# Supplementary Materials

Section S1 Device fabrication.
Section S2 High resolution STEM of heterostructure devices
Section S3 Atomic force microscopy
Section S4 Conductive atomic force microscopy
Section S5 Hysteresis of graphene and $MoS_2$ FETs
Section S6 Further examples of ReRAM memory elements with titanium adhesion layer
Section S7 Additional optoelectronic device data
Fig. S1 Heterostructure processing route.
Fig. S2 Additional TEM data.
Fig. S3 Atomic force microscopy data.
Fig. S4 Comparison of surface roughness of graphene on hBN and on $HfS_2$
Fig. S5 Conductive AFM on $HfO_x$.
Fig S6. Hysteresis behavior of graphene and $MoS_2$ FETs in different dielectric environments.
Fig S7 Comparison of hysteresis width ($\Delta V_H$) as a function of sweep rate for the hBN-$MoS_2$-$HfO_x$ and $SiO_2$-$MoS_2$-$HfO_x$ devices.
Fig. S8 Additional ReRAM memory devices.
Fig. S9 Temperature dependence of the resistance for a Graphite-$HfO_x$-Cr/Au vertical structure with t < 3 nm tunnel barriers.
Fig. S10 Additional optoelectronic characterization.
References (*60*, *61*)

**Acknowledgments:** We thank P. R. Wilkins and A. Woodgate for technical support. **Funding:** F.W acknowledges support from the Royal Academy of Engineering. J.D.M. acknowledges financial support from the Engineering and Physical Sciences Research Council (EPSRC) of the United Kingdom, via the EPSRC Centre for Doctoral Training in Metamaterials (Grant No. EP/L015331/1). S.R. and M.F.C. acknowledge financial support from EPSRC (Grant no. EP/K010050/1, EP/M001024/1, EP/M002438/1), from Royal Society international Exchanges Scheme 2016/R1, from The Leverhulme trust (grant title "Quantum Revolution" and "Quantum Drums"). A.P Rooney and S.J Haigh acknowledge support from the EPSRC postdoctoral fellowship and from the European Research Council (ERC) under the European Union's Horizon 2020 research and innovation programme (grant agreement ERC-2016-STG-EvoluTEM-715502) and the Defence Threat Reduction Agency (HDTRA1-12-1-0013). I.A. acknowledges financial support from The European Commission Marie Curie Individual Fellowships (Grant number 701704). **Author contributions:** N.P produced most experimental samples, contributed to measurements and contributed to writing the paper. J.M contributed to measurements and writing the paper. M.D.B contributed to sample production, measurements and writing the paper. A.D.S contributed to measurements, preparing figures and writing the paper. I.A provided AFM measurements, J.U.E and K.A produced ReRAM devices. A.P.R and S.J.H produced HRTEM sample and provided HRTEM data and contributed to writing the manuscript. S. R and M. F. C contributed to writing the manuscript. F.W conceived and supervised the project, contributed to sample production, measurements and writing of the paper. **Completing interests:** The authors declare that they have no competing interests. **Data and materials availability:** All data needed to evaluate the conclusions in the paper are present in the paper and/or the Supplementary Materials. Additional data related to this paper may be requested from the authors.

**Figures**

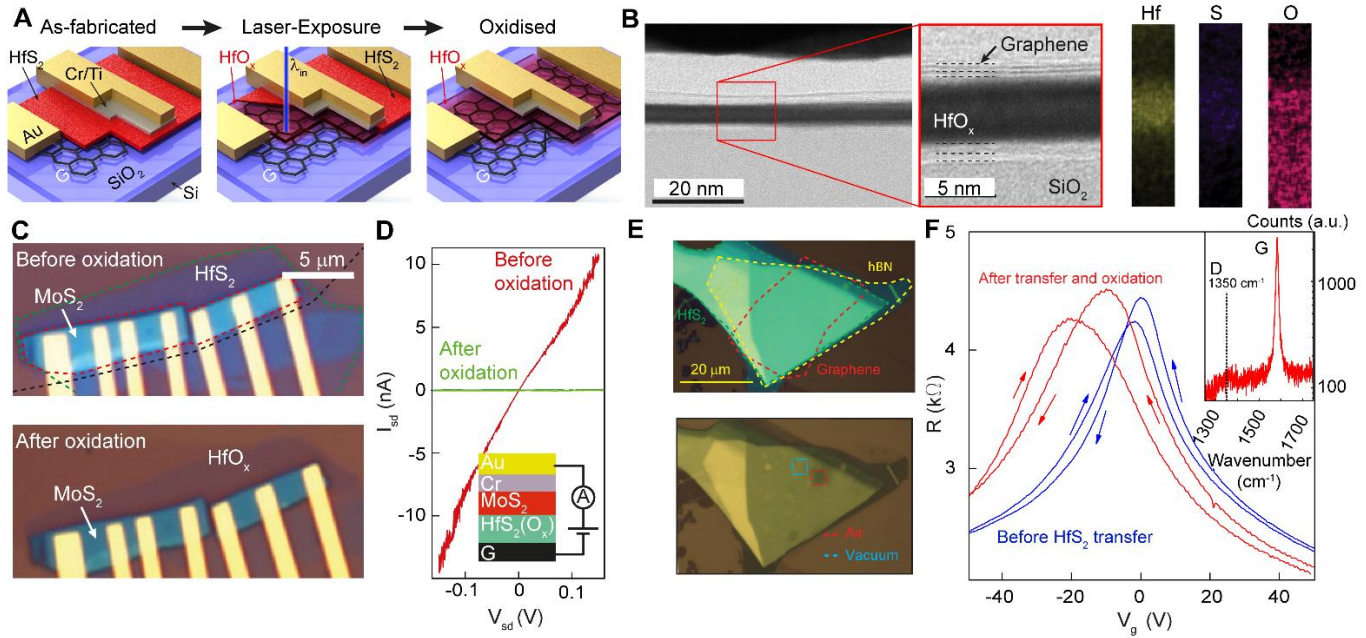

**Fig. 1. Heterostructure processing and characterization.** (**A**) The heterostructure is fabricated via dry transfer peeling from PDMS membrane (left), the area containing HfS$_2$ is exposed to laser light (center) and the HfS$_2$ is converted into HfO$_x$ (right). (**B**) BF STEM image showing a cross section of a Gr/HfO$_x$ device after laser-assisted oxidation (left) and EDX spectroscopy elemental analysis (right). (**C**) Optical image of a Graphene/HfS$_2$/MoS$_2$ heterostructure before (top) and after (bottom) oxidation. Black outlines the region of the graphene back gate, green -outlines the HfO$_2$ and red the MoS$_2$. (**D**) Current ($I_{sd}$) vs applied voltage ($V_{sd}$) for the heterostructure in panel **C** before (red) and after (green) photo-induced oxidation. Inset shows the stacking sequence. (**E**) (top) optical micrograph of a HfS$_2$ flake encapsulated between hexagonal boron nitride and graphene (green: HfS$_2$, yellow: hBN, red: graphene), (bottom) Optical micrograph of the same heterostructure imaged within our vacuum chamber showing laser irradiation effects in vacuum (blue hatched area) and in air (red hatched area). Note: no obvious oxidation effects are observed when irradiated in vacuum P ~ 10$^{-5}$ mbar. (**F**) Two-terminal resistance vs gate voltage for a graphene on hBN (d ~ 40 nm) /SiO$_2$ (290 nm) field effect transistor measured at T = 266 K in a helium atmosphere (blue curve) and after placing a thin HfS$_2$ flake and subjecting it to laser oxidation (red curve) (Sweep rate = 10 V/min). Inset shows a Raman spectrum of graphene after oxidation plotted on logarithmic scale showing the G peak and negligible D peak.

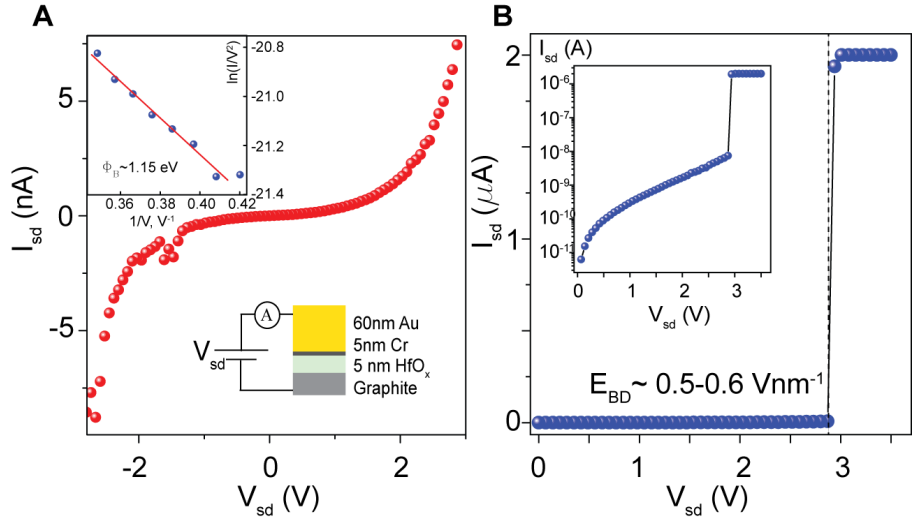

**Fig. 2. Break down of HfO$_x$ dielectrics.** (**A**) $I_{sd}$-$V_{sd}$ characteristics for a 5 nm Graphite/HfO$_x$/Cr/Au junction. Inset top left: Fowler-Nordheim tunneling theory, Bottom right: device schematic. (**B**) $I_{sd}$-$V_{sd}$ for an extended voltage range showing the breakdown field for the dielectric (inset: log scale plot of the same data showing the exponential dependence of tunneling current with bias voltage).

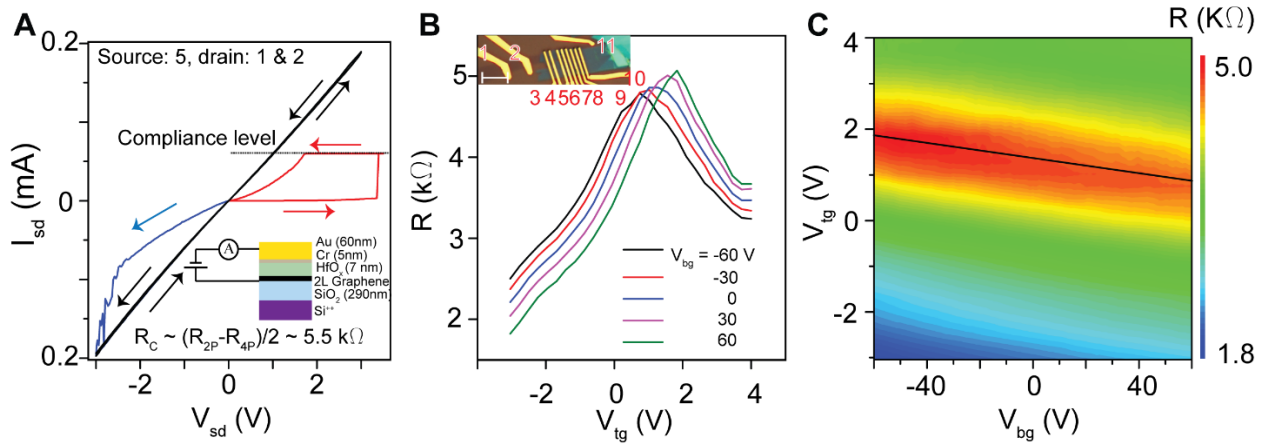

**Fig. 3. HfO$_x$ as an electrical contact material and gate oxide in a dual gated graphene FET.** (**A**) $I_{sd}$-$V_{sd}$ indicating the formation of a conductive filament in the oxide (red curve) further sweeps reduce resistance to 10's KOhm level (blue and black curve). (**B**) R($V_{tg}$) for different values of $V_{bg}$ from -60V to +60V. Inset: optical micrograph of the heterostructure device consisting of Gr-HfO$_x$-Cr/Au (scale bar: 10$\mu$m) . (**C**) Contour map of the channel resistance between contact 8-10 with contact 9 acting as the top gate electrode.

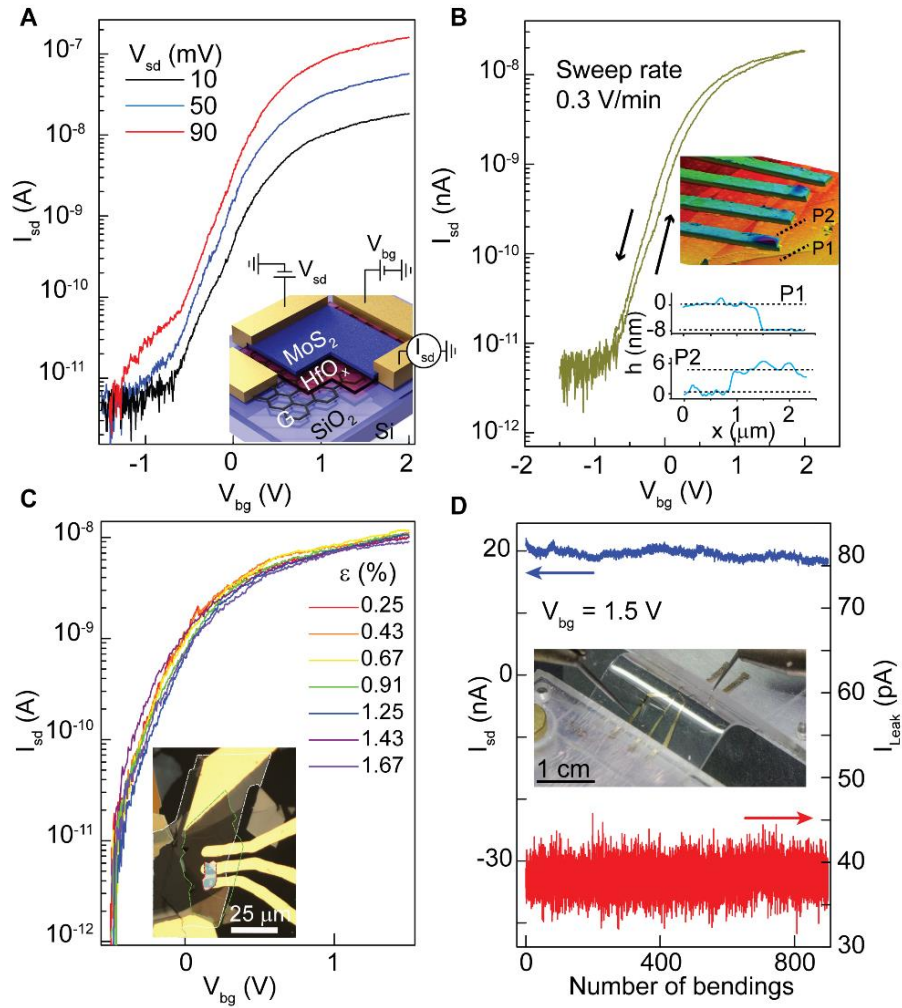

**Fig. 4. TMDC field effect transistors using photo-oxidized HfS$_2$** (**A**) $I_{sd} - V_g$ for a MoS$_2$ FET with an 8 nm oxidised HfS$_2$ film schematically shown in the inset. (**B**) forward and reverse sweeps highlighting small hysteresis. Inset: shows the AFM image for the device along with the height profile for the HfO$_x$ (P1) and the MoS$_2$ (P2). (**C**) Gate voltage dependence of the channel current for a MoS$_2$ FET on 0.5 mm PET substrate for different levels of strain up to 1.6 % at $V_{sd} = 10$ mV. Inset: optical micrograph of the device white highlight: graphite, green highlight: HfO$_x$ and red highlight: MoS$_2$. (**D**) Source-drain current at $V_{sd} = 20$ mV (blue) and gate leakage current (red) at $V_{gs} = 1.5$ V over 800 bending cycles.

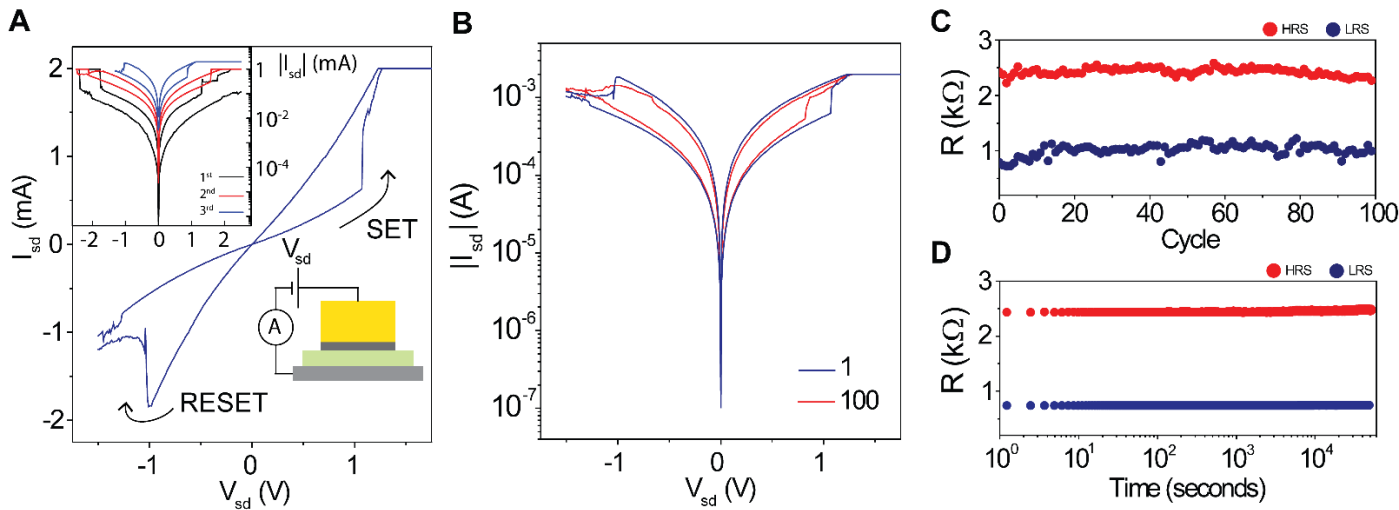

**Fig. 5. Example of a resistive switching memory element (ReRAM).** (**A**) Example of a switching cycle for the device architecture shown in the inset (bottom right). Top left inset: initial filament formation sweeps before repeatable switching was achieved. (**B**) 1st and 100th switching cycle for the same device (**C**) Resistance vs cycle number for the two resistance states plotted for the LRS (blue) and HRS (Red). (**D**) Time stability for the two resistance states.

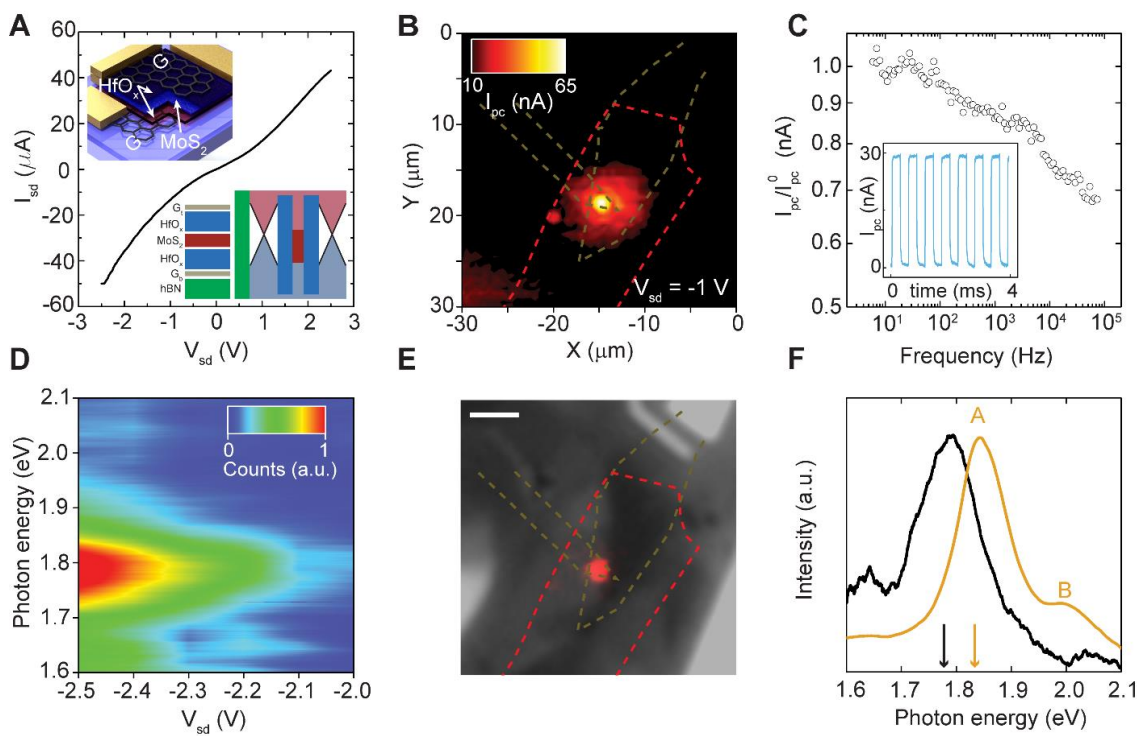

**Fig. 6. Thin HfO$_x$ barriers for optoelectronic applications.** (**A**) Current – Voltage characteristics for the single quantum well. Inset (top) illustration of the device architecture. Lower inset is schematic of the heterostructure band-alignment (hBN-Gr$_b$-HfO$_x$-MoS$_2$-HfO$_x$-Gr$_t$). (**B**) Scanning photocurrent map acquired with a bias of $V_{sd}$ = -1 V applied between the top and bottom graphene (**C**) Normalised photocurrent as a function of modulation frequency. Inset shows the temporal response of the photocurrent at f=1.8 kHz. (**D**) Colour map of the EL spectra as a function of $V_{sd}$. (**E**) False-color CCD image of the EL overlaid on an optical image of the device (Scale bar = 5 μm). (**F**) Comparison between the normalised intensities of the EL (black) and PL (brown) acquired under $V_{sd}$ = -2.5 V and $V_{sd}$ = 0V respectively.

# Supplementary Materials for

## *Laser writable high-K dielectric for van der Waals nano-electronics*

N. Peimyoo,[1] M. D. Barnes,[1] J. D. Mehew,[1] A. De Sanctis,[1] I. Amit,[1] J. Escolar,[1] K. Anastasiou,[1] A. P. Rooney,[2] S. J. Haigh,[2] S. Russo,[1] M. F. Craciun,[1] and F. Withers[1*]

Correspondence to: f.withers2@exeter.ac.uk

**This PDF file includes:**

# Section S1 Device fabrication

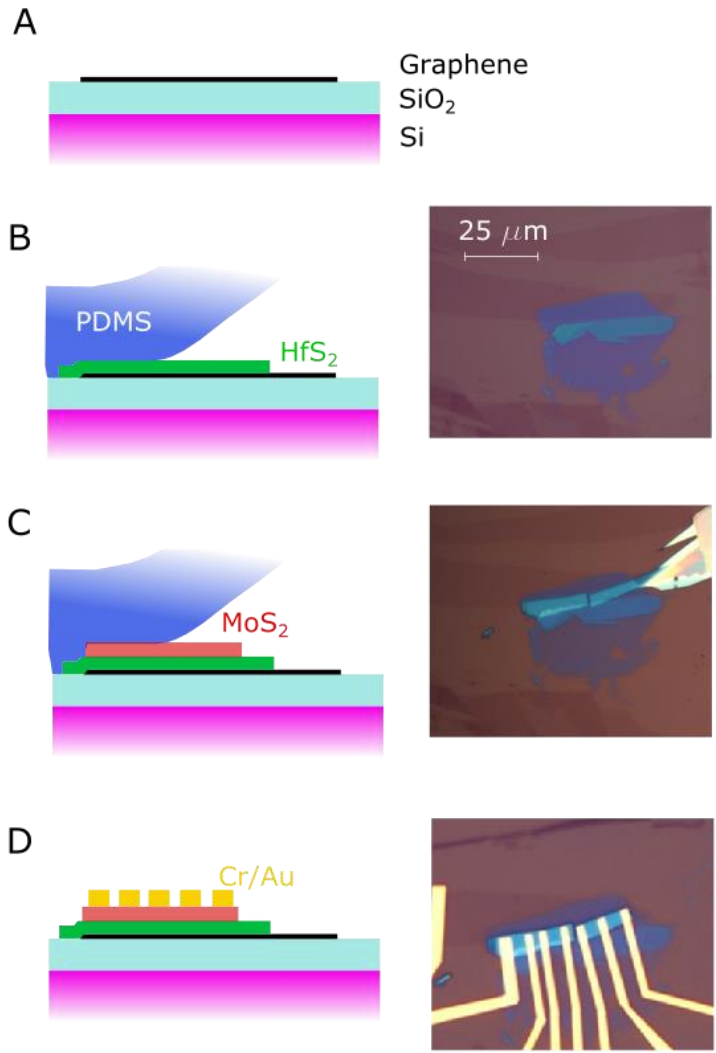

We make use of a PDMS stamp transfer technique. (*30*) Figure S1 shows a typical fabrication route for a $MoS_2$ FET. Firstly, graphene is mechanically exfoliated onto a thermally oxidized silicon wafer. After this the $HfS_2$ flakes are exfoliated onto PDMS and transferred to the graphene. The $HfS_2$ flakes are released from the PDMS between 50-60ºC. This process is then repeated for the subsequent layers of the device as shown in Figure S1C. After the heterostructure stack is formed conventional electron beam lithography is used to define electrical contacts.

The same process is used for other devices such as memory devices, dual gated graphene FET's and light emitting quantum well devices.

For devices on hBN substrates we use a PMMA membrane and dry peel the graphene from the PMMA onto the hBN (*25, 31*).

In this work the $HfS_2$ and $WSe_2$ was purchased from HQGraphene (http://www.hqgraphene.com/) whilst the $MoS_2$ and hBN crystals were acquired from Manchester Nanomaterials (http://mos2crystals.com/).

*Fig. S1. Heterostructure processing route.* (**A**) Graphene/graphite mechanically exfoliated onto a thermally oxidized silicon wafer with $SiO_2$ thickness of 290 nm (**B**) These $HfS_2$ flakes are transferred using a PDMS stamp onto the graphene (**C**) Graphene or TMDC's are then transferred by PDMS onto the $HfS_2$ layer. (**D**) Conventional mico-fabrication of Cr(5nm)/Au(50nm) contacts to the graphene/graphite back-gate and to the TMDC channel. Followed by plasma etching in $O_2$/Ar plasma.

# Section S2 High resolution STEM of heterostructure devices

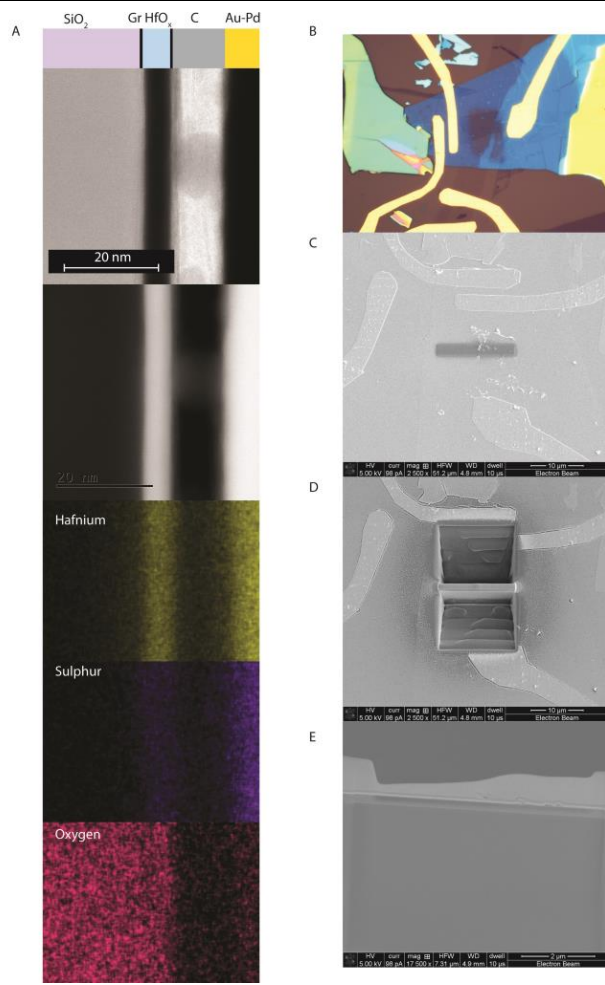

Figure S2 shows additional scanning transmission electron microscope (STEM) data for a cross-sectional device based on a Gr-HfO$_x$-Gr heterostructure The EDX spectroscopy elemental mapping confirms the structure and composition, see Fig S2A. Fig S2B is an optical image of the irradiated device imaged in A. Fig S2(C-E) shows scanning electron microsc (SEM) images illustrating the process of TEM sample preparation using FIB.

The details of cross-sectional STEM sample preparation can be found in previous reports (*9, 34*).

***Fig. S2. Additional TEM data***. (***A***) *Cross-section of a Graphene-HfO$_x$-Graphene heterostructure along with EDX spectroscopy elemental mapping of the heterostructure showing the HfO$_x$ layer and low sulphur content.* (***B***) *Optical micrograph of the device imaged in A, the central transparent region has been transformed into amorphous HfO$_x$ via laser irradiation.* (***C-E***) *SEM images of during FIB milling of the TEM sample.*

# Section 3 Atomic force microscopy

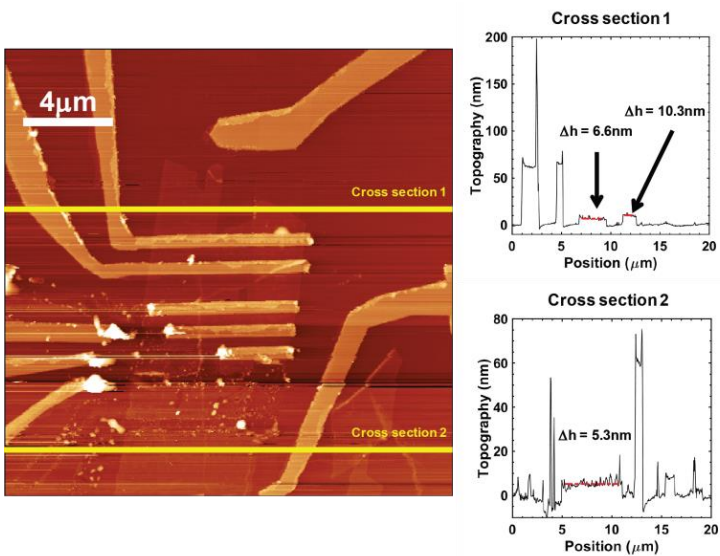

Tapping mode atomic force microscopy (AFM) was used to extract the flake thicknesses in different devices. Figure S3 shows the dual gated graphene FET shown in Figure 3 of the main text. The average thickness obtained from several cross-sections was 7.4 nm.

AFM can also be used to understand how clean the interface is formed between two materials. Figure S4 shows a hBN-$HfS_2$-Graphene heterostructure. We find that the roughness of graphene on $HfS_2$ is comparable to that of graphene on hBN.

*Fig. S3. Atomic force microscopy data* (Left) AFM topography of the dual gated bilayer graphene FET shown in Figure 3 of the main text. (Right) cross-sections of the height profile for the $HfO_x$ dielectric.

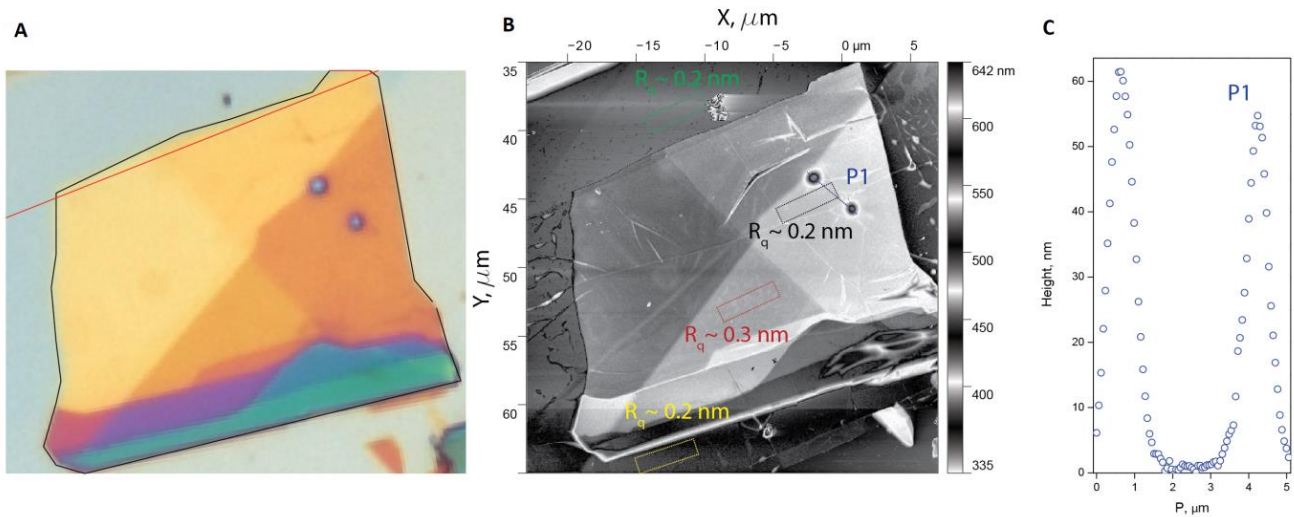

*Fig. S4. Comparison of surface roughness of graphene on hBN and on $HfS_2$* (A) Optical image of the hBN-$HfS_2$-Graphene heterostructure stack. (B) Tapping mode atomic force microscopy of a graphene encapsulated $hfS_2$ flake (note, the RMS roughness, $R_q$ of the hBN (green), graphene on hBN (yellow) and graphene on $HfS_2$ (black and red) are of the same order ~ (0.2-0.3) nm)  and (C) height profile of bubbles of contamination trapped between the hBN-$HfS_2$ interface, marked P1 in (B).

# Section S4 Conductive atomic force microscopy

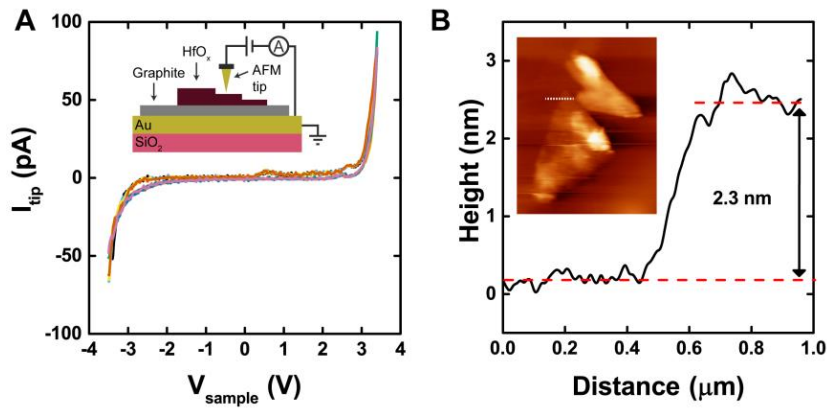

***Fig. S5 Conductive AFM on HfOx**. (**A**) Multiple IV curves acquired on flake by applying a voltage to the sample and measuring the current with the AFM tip. Inset shows measurement schematic. (**B**) Height profile of flake. Inset is topographical image of sample*

Conductive AFM was used to locally probe the tunneling current through ultrathin $HfO_x$. A voltage was applied to the graphite/Au substrate and the current measured using a conductive tip (diamond like carbon) connected to a Femto DLPCA current amplifier and voltmeter. Figure S5 shows multiple I-V curves for a 2.3 nm thick flake of $HfO_x$. Topographical image analysis and height profile extraction were performed with WSxM v9.1 software.(*60*)

# Section S5 Hysteresis of graphene and MoS₂ FETs

To further investigate the quality of our photo-oxidized HfO$_x$, we fabricate three different heterostructure FETs (i) hBN-graphene-HfO$_x$ (ii) hBN-MoS$_2$-HfO$_x$ (iii) SiO$_2$-MoS$_2$-HfO$_x$ and the hysteresis of these devices are measured with different sweep rates (Figure S6). We find large hysteresis in SiO$_2$-MoS$_2$-HfO$_x$ device and the hysteresis width ($\Delta V_H$) increases significantly, whilst $\Delta V_H$ is reduced for MoS$_2$ encapsulated between hBN and HfO$_x$. An increase in hysteresis of MoS$_2$ in contact with SiO$_2$ is consistent with previous reports, which is originated from charge traps at the interface between MoS$_2$ and SiO$_2$. We also observe a negligible level of hysteresis for a graphene transistor encapsulated between hBN and HfO$_x$.

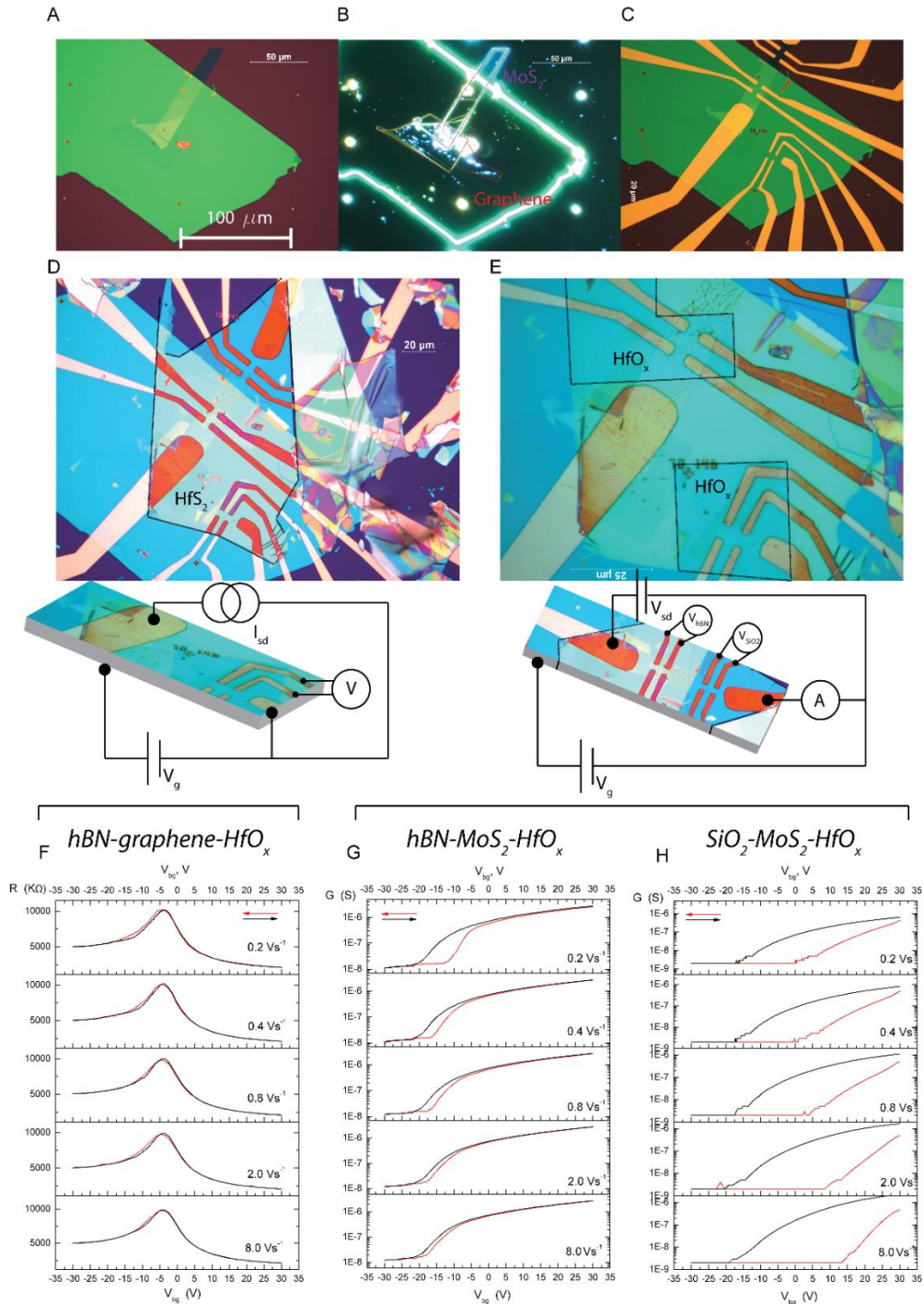

*Figure S6. Hysteresis behavior of graphene and MoS₂ FETs in different dielectric environments.* (A) Optical image of heterostructures consisting of different stacking sequences. (B) Dark-field image of (A) showing the overlapped regions and the outlines of each material (B). (C) The fabrication of Cr/Au contacts on Hall-bar geometry graphene and MoS$_2$ channels. (D) Optical image of the devices in (C) encapsulated by a large HfS$_2$ flake (black highlight). (E) Optical image of the corresponding device after

Page **20** of 24

*laser irradiation in the regions marked as HfOₓ. The final device configurations include (i) hBN-graphene-HfOₓ (ii) hBN-MoS₂-HfOₓ (iii) SiO₂-MoS₂-HfOₓ. (F) Resistance vs backgate voltage for graphene encapsulated between HfOₓ and hBN for different sweep rates. (G) Conductance vs backgate voltage for hBN-MoS₂-HfOₓ. (H) and SiO₂-MoS₂-HfOₓ (H) measured with different sweep rates. (F-H) top: Device measurement schematics. Arrows indicate the gate voltage sweep direction.*

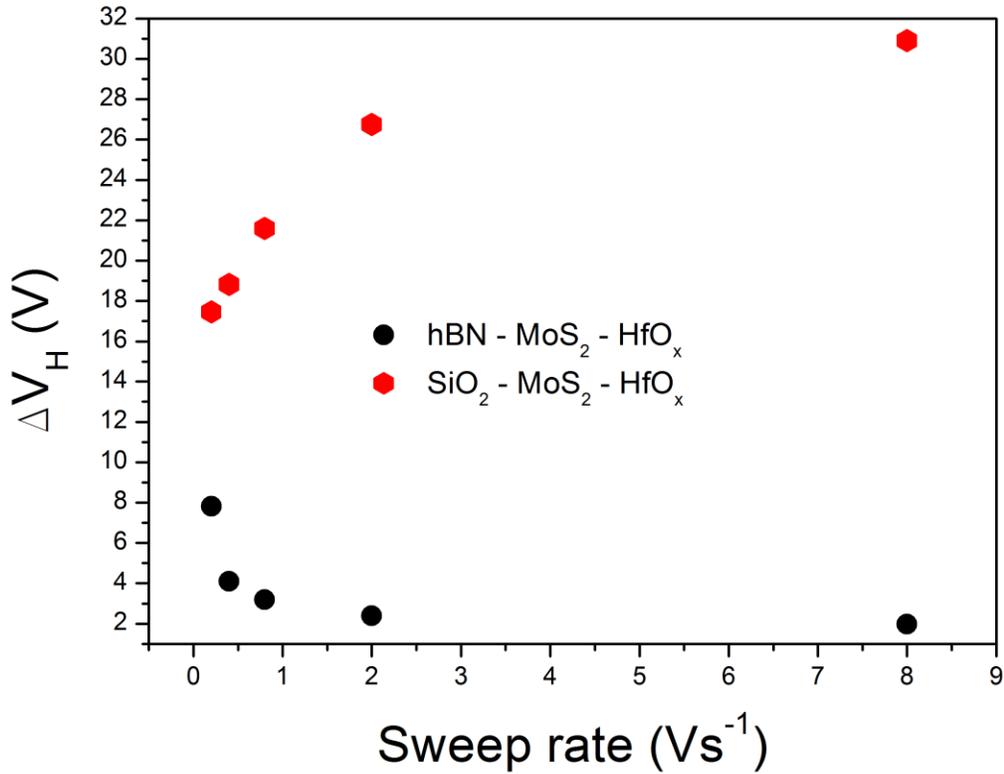

*Figure S7. Comparison of hysteresis width ($\Delta V_H$) as a function of sweep rate in the hBN-MoS₂-HfOₓ and SiO₂-MoS₂-HfOₓ devices.*

# Section S6 Further examples of ReRAM memory elements with titanium adhesion layer

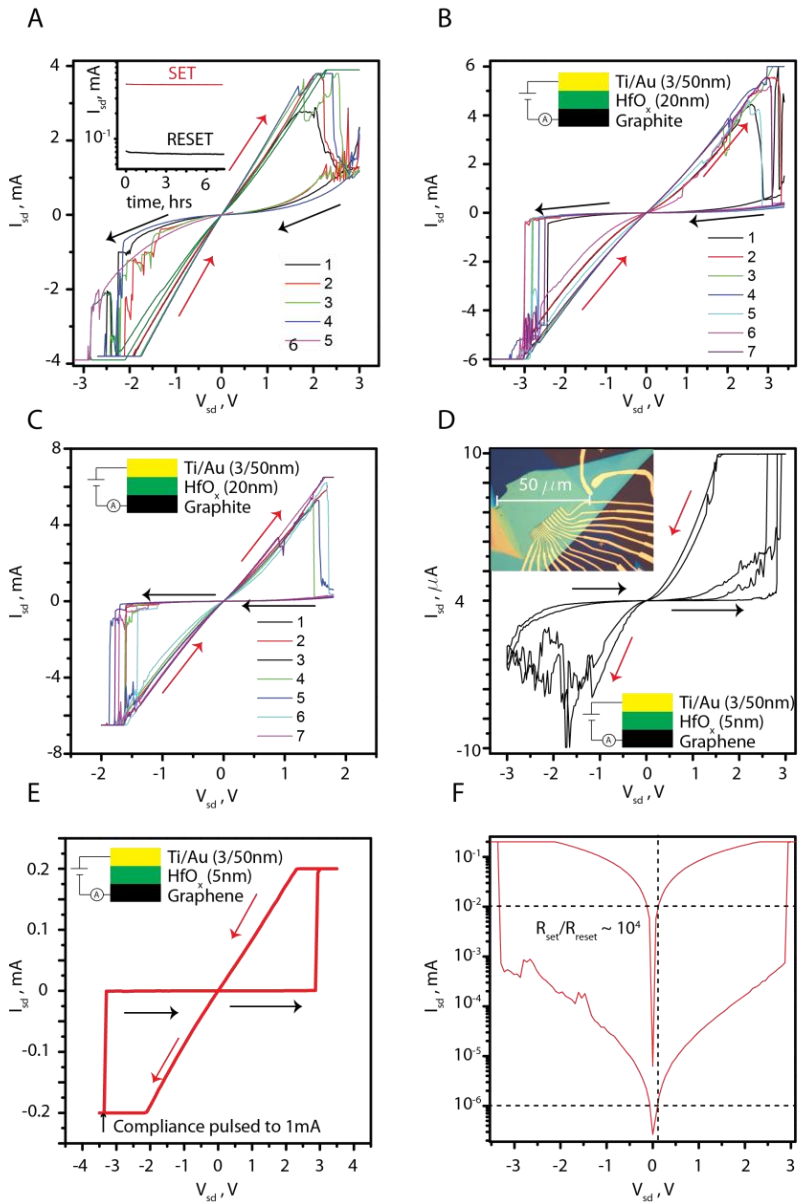

Figure S8 shows further examples of ReRAM heterostructure devices based on laser oxidised $HfO_x$ barrier material.

Figure S8 (A-C) shows resistive switching properties for a 20 nm thick $HfO_x$ barrier at different compliance levels from 4 mA to 6 mA.

While Figure S8D shows an example of a 5nm $HfO_x$ barrier.

Figure S8E shows an example of a 5nm resistive switching element which displayed a large $R_{SET}/R_{RESET} = 10^4$. In this device the RESET compliance level was pulsed to 1 mA to break the filament.

*Fig. S8. Additional ReRAM memory devices. (A-C) Resistive switching data for several cycles for a heterostructure consisting of graphite-$HfO_x$-Ti/Au for increasing current compliance levels. (D) Example of resistive switching for a thinner 5nm $HfO_x$ barrier. (E-F) Resistive switching device displaying large $R_{SET}/R_{RESET}$.*

# Section S7 Additional optoelectronic device data

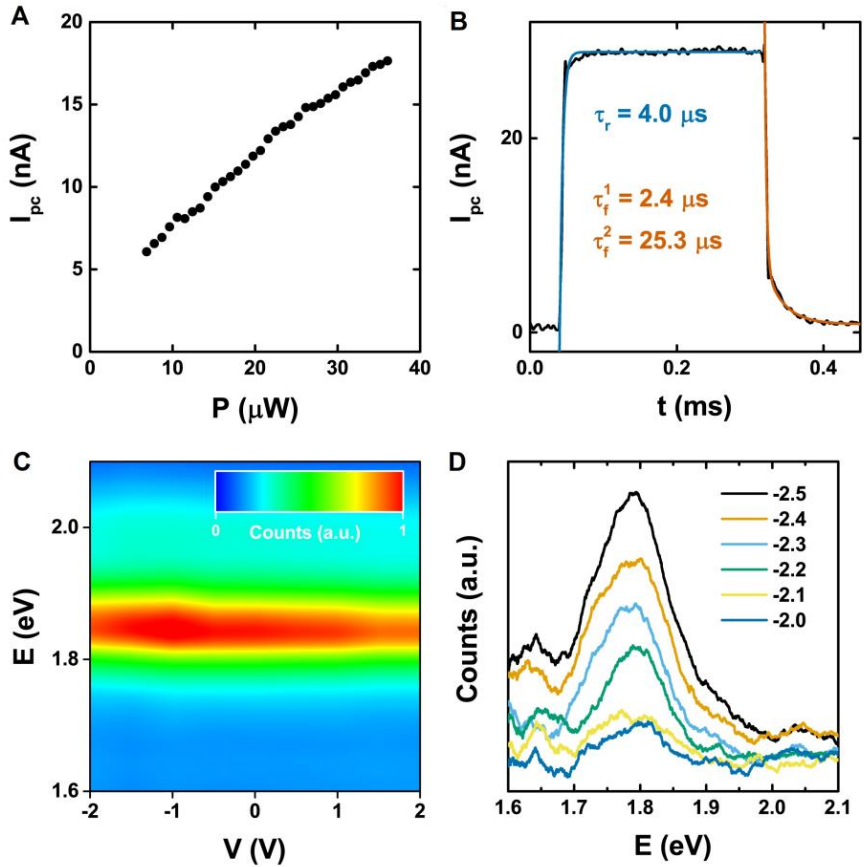

***Fig. S10. Additional optoelectronic characterization.*** *(A) Power dependence of the generated photo-current recorded at $V_{sd}$ = -1 V. (B) Single oscillation of the photo-current. The rise (fall) times have been extracted by fitting the data, black line, with exponential decays with one (two) time constants. (C) PL intensity for different applied bias voltages. (D) EL spectra for different bias voltages extracted from the contour map shown in Fig 6D of the main text.*

In the Figure 6, we identified the active area of the $HfO_x$ QW through photocurrent mapping. A hotspot in the photocurrent is seen which corresponds to region in which the graphene electrodes overlap the $HfO_x$ encapsulated $MoS_2$ flake. This localized photocurrent (Ipc), Figure S10A, has a non-linear power dependence which has been previously attributed to absorption saturation or electric field screening by the photoexcited carriers in $MoS_2$ (*61*). We observe an external quantum efficiency (EQE) of ($\eta = \frac{I_{pc}}{q}\frac{h\nu}{P} \sim 0.002\,\%$) smaller than previous works (*61*) which we anticipated due to the low absorption of monolayer $MoS_2$, the increased confinement of charges in the $HfO_x$ QW, and off-resonance excitation. Furthermore, the low EQE demonstrates that there is no significant gain mechanism present in our device. This corroborates with the rise and fall time analysis, Figure S10B, which reveals an exponential decay with two-time constants, similar in magnitude. Therefore, it is unlikely that one results from long lived charge trapping – a common mechanism for photoconductive gain. As a result, this conclusion supports our claim of the formation of a clean oxide with few impurity states, crucial for the creation of a quantum well.

Aside from photocurrent generation, the extraction of carriers also manifests as a bias dependence of the photoluminescence (PL). For positive bias voltages the PL intensity decreases to a minimum at 2 V. As we increase the bias the Fermi level of the bottom graphene electrode aligns with the conduction band of $MoS_2$ favoring the extraction of photoexcited carriers preventing their recombination and quenching the PL. Similarly, as we sweep the bias to negative values we observe first a peak in PL intensity followed by a decrease with the peak located away from zero likely due to asymmetry in the thickness of the two barriers and the doping of top and bottom graphene.

Upon increasing the bias to more negative values (V < -2 V) we begin to observe electroluminescence (EL) as seen in Figure 6E and discussed in the main text. In Figure S10D we present EL spectra taken at 0.1 V intervals showing the emergence of the main peak at 1.8 eV.

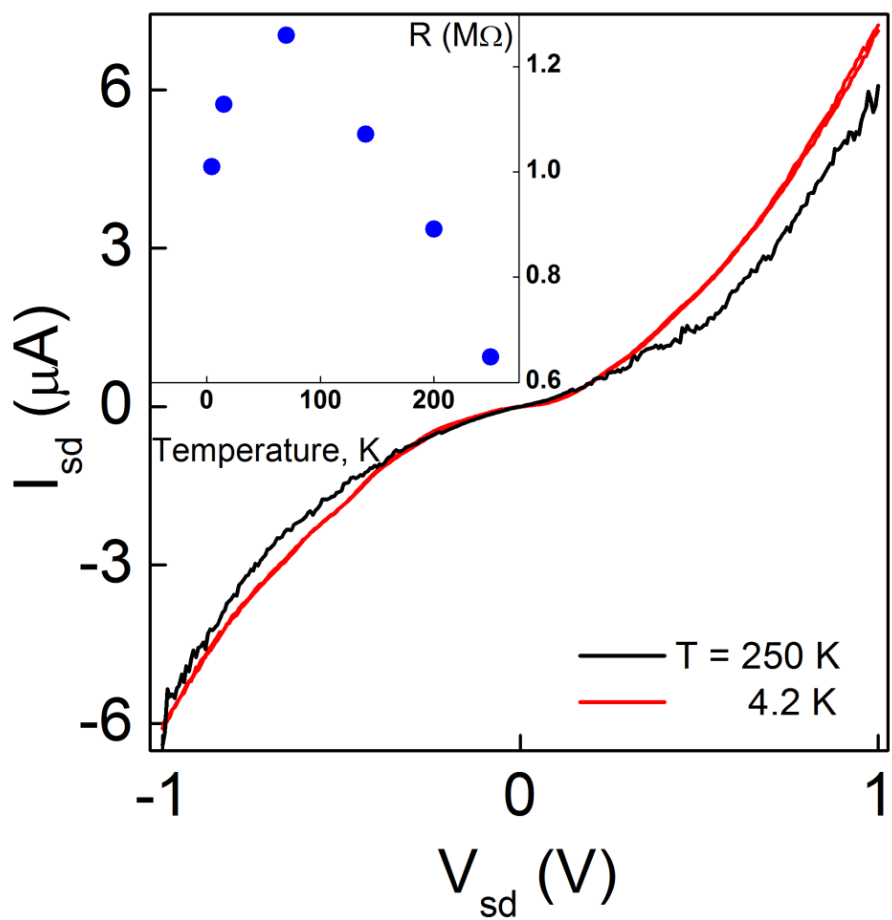

***Fig. S9. Temperature dependence of the resistance for a Graphite-HfO$_x$-Cr/Au vertical structure with t < 3 nm tunnel barriers.*** $I_{sd}$-$V_{sd}$ *characteristics measured at T = 4.2 and 300 K (Junction area: 8 mm$^2$). Inset: extracted low bias junction resistance vs temperature.*